\documentstyle[epsf,pra,aps,preprint]{revtex}

\begin{document}
\draft

\def\del{\partial}
\def\be{\begin{equation}}
\def\ee{\end{equation}}
\def\bea{\begin{eqnarray}}
\def\eea{\end{eqnarray}}
\def\nn{\nonumber}
\def\en{\begin{enumerate}}
\def\een{\end{enumerate}}
\def\ra{\rightarrow}

\def\lsim{\hbox{\lower .8ex\hbox{$\, \buildrel < \over \sim\,$}}}
\def\gsim{\hbox{\lower .8ex\hbox{$\, \buildrel > \over \sim\,$}}}

\title{Discrete Thermodynamic Bethe Ansatz}
\author{Michel Berg\`ere$^+$,
Ken-Ichiro Imura$^{++}$ and St\'ephane Ouvry$^{++}$}

\address{$^+$ Spht,  CEA-Saclay }

\address{$^{++}$ LPTMS, Universit\'e Paris-Sud}

\address{bergere@spht.saclay.cea.fr; imura@ipno.in2p3.fr; ouvry@ipno.in2p3.fr}

\date{\today}
\maketitle

\begin{abstract}

 We propose discrete TBA equations for models with discrete spectrum. We
illustrate our construction on the Calogero-Moser model
and determine the  discrete 2-body TBA function
which yields the exact $N$-body Calogero-Moser thermodynamics. We 
apply
this algorithm  to the Lieb-Liniger model in a harmonic well, a model
which is relevant
for the microscopic description of harmonically trapped Bose-Einstein
condensates in one dimension.
 We find  that the discrete TBA reproduces correctly the $N$-body groundstate 
 energy of the Lieb-Liniger model in a harmonic well
 at first order in perturbation theory,  but corrections do 
 appear at 
second order. 
\end{abstract}


\section{Introduction}

It has been known for a long time that the spectrum of the 
Calogero model \cite{calogero},
defined here as particles  on an infinite line interacting via
$\alpha(\alpha-1)/(x_i-x_j)^2$ 2-body interactions, can be found \cite{poly} by
the Bethe Ansatz (BA) which assumes periodic boundary conditions,
and that its thermodynamics 
can be obtained,  in the thermodynamic limit,  by
the Thermodynamic Bethe Ansatz (TBA) \cite{YY}.
 It is however not necessary   to rely on the  BA (or the TBA)
to get the 
Calogero spectrum (or its thermodynamics). Indeed,
the Calogero model is
 exactly solvable

- either by confining the particles in a harmonic well of frequency $\omega$:
the Calogero-Moser model \cite {CM} 
with 
discretized harmonic well quantum numbers and energies, and Hamiltonian

\be H_N=-{1\over 2}\sum_{i=1}^N{d^2\over dx^2_i}+\alpha(\alpha-1)
\sum_{i<j}{1\over (x_i-x_j)^2}+{1\over 2}\omega^2x_i^2\ee

- or by confining the particles in a periodic box of length $L$: 
the Calogero-Sutherland model \cite{sutherland1}
 with discretized momenta and energies,
and Hamiltonian
\be H_N=-{1\over 2}\sum_{i=1}^N{d^2\over dx^2_i}+\alpha(\alpha-1)
({\pi\over L})^2\sum_{i<j}{1\over \sin^2[{\pi\over L}(x_i-x_j)]}\ee
(the $1/\sin^2[\pi(x_i-x_j)/L]$  interactions
are nothing but the periodic version of the infinite
line interactions). 
It is not a surprise that Bethe ansatz equations yield
the Calogero-Sutherland spectrum, since they  also assume, as stressed above, 
periodic boundary conditions.

Both parameters $\omega$ and $L$ can be considered
as long distance regulators, the thermodynamic limit, i.e. the
infinite line limit, being obtained either
by  $\omega\to 0$ or $L\to\infty$, resulting in  continuous momenta and  
energies.

The Calogero 
model describes particules with intermediate  statistics,
which is natural due to the topological (statistical) nature of the $1/(x_i-x_j)^2$ interaction
in 1d.  
In the thermodynamic limit indeed \cite{isakov1}, the Calogero thermodynamics  realizes
microscopically
Haldane (Hilbert space counting) statistics \cite{haldane}. 
Moreover, the Calogero model has been shown to be obtained as  the
vanishing magnetic field limit \cite{ouvry1} of the 
lowest Landau level anyon model \cite{ouvry2}
(LLL-anyon model) in the regime where the flux tubes carried by the anyons
screen the flux of the external magnetic field (screening regime). Not
surprisingly,
the
LLL-anyon model  also realizes  microscopically   Haldane statistics,
the Hilbert space
counting argument being manifest here via a mean field argument (adding
anyons screen the external magnetic field, and thus diminish the Landau
degeneracy of the total -mean+external- magnetic field):  
thus a clear relation between 
Haldane \cite{wu} and anyon statistics \cite{LM}.

Starting from  the BA spectrum, and  following Yang and Yang
footsteps \cite{YY}, one can compute  \`a la TBA
the thermodynamics of the Calogero model in the
thermodynamic limit  $L\to \infty$.  The thermodynamical
potential $\ln Z$ -where $Z=\sum_{N=0}^\infty z^{N}Z_N$
is the grand partition function- ends up to be
those of a
Fermi gas\footnote{There is an equivalent formulation in terms of a free
Bose gas, namely
$$
\log Z=
{L\over 2\pi}\int_{-\infty}^\infty dk 
\log{1\over 1-ze^{-\beta{\tilde{\epsilon}}({k})}},
$$
 but with a 1-body  energy ${\tilde{\epsilon}}(k)$ defined  as
$$
\beta{\tilde{\epsilon}}({k_1})=\beta \epsilon_o(k_1)
-{L\over 2\pi}\int_{-\infty}^\infty dk_2
{\tilde{\Phi}}(k_1-k_2)
\log{1\over 1-ze^{-\beta{\tilde{\epsilon}}({k_2})}}
$$
One has obviously
$$
{\tilde{\Phi}}(k_1-k_2)-{{\Phi}}(k_1-k_2)=-{2\pi\over L}\delta(k_1-k_2)
$$} 
\be\label{1}
\log Z=
{L\over 2\pi}\int_{-\infty}^\infty dk 
\log[1+ze^{-\beta\epsilon({k})}],
\ee
 but with a 1-body  energy $\epsilon(k)$ defined in terms
of the free continuous 1-body quadratic spectrum $\epsilon_o(k)=k^2/2$ as 

\be\label{2}
\beta\epsilon({k_1})=\beta \epsilon_o(k_1)
-{L\over 2\pi}\int_{-\infty}^\infty dk_2
\Phi(k_1-k_2)
\log[1+ze^{-\beta\epsilon({k_2})}]
\ee

In the Calogero case \cite{BW},  
\be \label{3}\Phi(k_1-k_2)={2\pi\over L}(1-\alpha)\delta(k_1-k_2)\ee
is intimately
related  to the 2-body scattering angle, and  encodes,  
if one  thinks in terms
of statistics,  the statistical exclusion between two quantum states, here   
with  the same momentum.

Note that if one denotes $y(k)=1+ze^{-\beta\epsilon({k})}$, which  can be
regarded, in view of (\ref{1}), as the
grand partition function at momentum $k$, then 
(\ref{2}) can be rewitten  as 
\be\label{4}  y(k)-ze^{-\beta{\epsilon_o(k)}}y(k)^{1-\alpha}=1\ee
a particular case of Ramanujan equations \cite{rama}.

As already said, equations of the type (\ref{1},\ref{4}) were first
obtained directly by
i) considering 
the exact
$N$-body Calogero spectrum in a harmonic well \cite{isakov1}, or   by considering the exact $N$-body LLL-anyon spectrum in a 
harmonic well \cite{ouvry2},  ii) and then
taking the thermodynamic limit.

Now we might ask the following question: are  the TBA equations (\ref{1},\ref{2})
specific to the thermodynamic
limit with continuous momenta and continuous dressed energies, or can they 
also describe the thermodynamic of the Calogero model in
a harmonic well or in a periodic box with discretized energies?
In other words, can we find  a discretized version of the function
$\Phi(k_1-k_2)$
in (\ref{3}) such that the
harmonic well or periodic box
Calogero thermodynamics
narrow down to a set of defining equations analogous to (\ref{1},\ref{2})?

We will show  that  the thermodynamic of the Calogero model in a
harmonic well can  indeed
be rewritten ``\`a la  TBA''  in terms of a discretized function
$\Phi$  
which will
encode  the statistical Calogero exclusion between   different
discrete harmonic  energy levels
and which will, as it should, reproduce, in the thermodynamic limit 
$\omega\to
0$, $\Phi(k_1-k_2)$ in (\ref{3}). 
Not surprisingly, the same conclusion will be reached for the LLL-anyon 
model in an harmonic
well, whose thermodynamics will obey the same TBA equations as the
Calogero-Moser thermodynamics.

We will also argue that the same logic  applies in the Calogero-Sutherland 
case,  provided that a global  shift of the bare quantum numbers is made
in order to maintain a symmetric repartition of the dressed quantum numbers
 around zero.

 Finally, we will look at possible 
applications of   discrete TBA thermodynamics
 beyond the Calogero-Moser and harmonic LLL-anyon cases, by
considering
the Lieb-Liniger model in a harmonic well
\cite{LL},\cite{myrheim}. This model
is interesting because of its relevance to the description of one
dimensional trapped Bose condensates \cite{BE}.
We will show  that  the $N$-body groundstate energy is correctly
reproduced at first order in perturbation theory by the discrete TBA equations,
but corrections do appear at 
second order.

\section{The Calogero case}

For a system with a discrete 1-body harmonic spectrum, the TBA equations 
 (\ref{1},\ref{2}) should
rewrite quite generally as 
\be\label{5}
\log Z=\sum_{n=0}^\infty \log[1+z e^{-\beta\epsilon({n})}],
\ee
where 
the 1-body dressed
energy $\epsilon(n)$ should now be defined in terms of the 1-body  1d harmonic
spectrum (bare spectrum) $\epsilon_o(n)=(n+1/2)\omega,\quad 0\le n$ as 
\be\label{6}
\beta\epsilon({n_1})=\beta\epsilon_o(n_1)-\sum_{n_2=0}^{\infty}\Phi_{n_1,n_2}
\log[1+z e^{-\beta\epsilon({n_2})}]
\ee
(\ref{5},\ref{6}) are just the discretized versions of (\ref{1},\ref{2}).
In (\ref{6}),
$ \Phi_{n_1,n_2}$ has  to be understood  as
 acting  on the free harmonic spectrum, i.e as acting on the
power series in $z$ obtained from (\ref{6})
by expanding $y(n)=1+ze^{-\beta\epsilon({n})}$ 
as
\be\label{7}
y(n)=\sum_{N=0}^\infty y_{N}(n)z^{N}
\ee
The lowest order terms of (\ref{7}) are
\bea\label{8}
y_0(n_1)&=&1,
\nn \\
y_1(n_1)&=&e^{-\beta\epsilon_o(n_1)},
\nn \\
y_2(n_1)&=&\sum_{{n_2}=0}^{\infty}
e^{-\beta\epsilon_o(n_1)}
\Phi_{n_1,n_2}
e^{-\beta\epsilon_o(n_2)},
\nn \\
y_3(n_1)&=&
\sum_{{n_2,n_3}=0}^{\infty}
e^{-\beta\epsilon_o(n_1)}
\left(
\Phi_{n_1,n_2}
e^{-\beta\epsilon_o(n_2)}
\Phi_{n_2,n_3}
+{1\over 2}
\Phi_{n_1,n_2}
e^{-\beta\epsilon_o(n_2)}
\Phi_{n_1,n_3}
\right)
e^{-\beta\epsilon_o(n_3)}
\nn \\
&-&
{1\over 2}\sum_{n_2=0}^{\infty}
e^{-\beta\epsilon_o(n_1)}
\Phi_{n_1,n_2}
e^{-2\beta\epsilon_o(n_2)},
\eea
where $0\le n_1,n_2,\cdots$ and the  summation
should be taken for all possible independant  integers 
$n_2,n_3,\cdots$.

If the TBA cluster coefficients
obtained from expanding $\log Z=\sum b_nz^n$
\bea
 b_1&=&\sum_{n=0}^{\infty}y_1(n),
\nn \\
b_2&=&\sum_{n=0}^{\infty}\left[y_2(n)-{1\over 2}y_1^2(n)\right],
\nn \\
 b_3&=&\sum_{n=0}^{\infty}
\left[
y_3(n)-y_1(n)y_2(n)+{1\over 3}y_1^3(n)
\right],
 \eea
have to match against
the  Calogero-Moser cluster coefficients
$ b_1={e^{-{\beta\omega\over 2}}\over 1-e^{-\beta\omega}}$,
$ b_2= 
e^{-\beta\omega}
\left[
{e^{-\beta\omega\alpha}-e^{-\beta\omega}\over
(1-e^{-2\beta\omega})(1-e^{-\beta\omega})}
-{1\over 2}{1\over 1-e^{-2\beta\omega}}
\right]$, $\ldots $,
 $\Phi_{n_1,n_2}$ should be defined as 
\be\label{9}
\Phi_{n_1,n_2}=P_{n_1,n_2}(\alpha)-P_{n_1,n_2}(\alpha=1)
\ee
where $P_{n_1,n_2}(\alpha)$ projects the two independant quantum numbers $0\le
n_1,n_2$ on
  dressed quantum numbers which, not surprisingly, obey
 exclusion statistics.  More precisely, 
 evaluating in
  (\ref{5},\ref{6},\ref{8}) expressions of the type  
\be 
\sum _{n_2=0}^{\infty}P_{n_1,n_2}(\alpha)e^{-\beta\epsilon(n_2)},\ee
$P_{n_1,n_2}(\alpha)$ amounts, $n_1$ being given,
 to the shift
 \be n_2\to
n_1+\tilde{n}_2+\alpha, \quad \tilde{n}_2\ge 0\ee
and  the summation over $n_2$ is replaced by the summation over 
${\tilde{n}}_2\ge 0$. In other words,
in terms of the independant quantum numbers  $0\le n_1, n_2$, 
denoting ${\bf n}_1=n_1, {\bf n_2}=n_1+\tilde{n}_2$,
$P_{n_1,n_2}(\alpha)$ means $ n_1\to {\bf {n}}_1$,
and $ n_2\to
{\bf {n}}_2+\alpha$, where $0\le {\bf {n}}_1\le {\bf {n}}_2$
are now bosonic quantum numbers.
Therefore  $P_{n_1,n_2}(0)$  projects
$0\le n_1,n_2$
onto  bosonic quantum numbers $0\le {\bf{n}}_1\le {\bf{n}}_2$,  whereas
$P_{n_1,n_2}(1)$ projects $0\le n_1,n_2$ onto fermionic quantum numbers.
Note that  in (\ref{9}) substracting $ P_{n_1,n_2}(\alpha=1)$
is simply a matter of
 convention, i.e. as stressed above, a fermionic
thermodynamical potential (\ref{1}) with a  spectrum
which has to coincide
with the bare spectrum when $\alpha=1$ -the Bose convention would yield
${\tilde{\Phi}}_{n_1,n_2}= P_{n_1,n_2}(\alpha)- P_{n_1,n_2}(\alpha=0)$.

More generally  notice that (\ref{9}) allows to rewrite (\ref{6}) as
\be\label{11}
y(n_1)-ze^{-\beta\epsilon_o(n_1)}
{\prod_{\tilde{n}_2=
0}^{\infty}y(n_1+\tilde{n}_2+\alpha)\over\prod_{\tilde{n}_2=
0}^{\infty}y(n_1+\tilde{n}_2+1)}=1
\ee
which can be viewed as the discretized version of
(\ref{4}).

Going one step further one gets
\be\label{12}
\prod_{\tilde{n}_2= 0}^{\infty}y(n_1+\tilde{n}_2)=\prod_{\tilde{n}_2= 0}^{\infty}
y(n_1+\tilde{n}_2+1)
+
ze^{-\beta\epsilon_o(n_1)}
\prod_{\tilde{n}_2= 0}^{\infty}y(n_1+\tilde{n}_2+\alpha)
\ee
which in turn, taken  at $n_1=0$, rewrites as
\be\label{13} 
Z=ze^{-\beta\epsilon_o(0)}
\prod_{\tilde{n}_2=0}^{\infty}y(\tilde{n}_2+\alpha)
+\prod_{\tilde{n}_2=0}^{\infty}y(\tilde{n}_2+1)
\ee

Equation (\ref{13}) can be interpreted by saying that either the first
particle is in the
groundstate at energy ${1\over 2}\omega$ and  then the next particle  is in
the energy
level at least higher than $({1\over 2}+\alpha)\omega$, or the
groundstate
is vacant and the first particle  
is in the  energy level at least higher than $({1\over 2}+1)\omega$.
One  verifies recursively that $Z$  in
(\ref{13}) is identical to the grand
partition of the Calogero-Moser model with $N$-body spectrum 
\be \label{15} E_N=\omega\sum_{i=1}^N\left([{\bf n}_i+\alpha (i-1)]+{1\over 2}\right)\ee
where $0\le {\bf n}_1\le {\bf n}_2\le \ldots $. 	
In terms of the bare independent
quantum numbers $0\le n_1,n_2,\ldots $, one has
$n_i\to {\bf n}_i+\alpha(i-1)=n_{i-1}+\tilde{n}_i+\alpha(i-1)$ with $
\tilde{n}_i\ge 0$. This is indeed a BA
``like''
spectrum, i.e. in terms of the dressed quantum numbers
${\bf n}'_i={\bf n}_i+\alpha(i-1)$, one has
${\bf n'}_i= {\bf n_i}+\alpha\sum_{j\ne i}\theta({\bf n'}_i-{\bf
n'}_j)$. In the 2-body case, it indeed amounts to $n_1\to n_1, n_2\to
n_1+\tilde{n}_2+\alpha$, i.e. to the the action of the
projector $P_{n_1,n_2}(\alpha)$ on the independent quantum
numbers $0\le n_1,n_2$.

Note that (\ref{9}) implies that the $N$-body
partition function $Z_N$ obtained from (\ref{5}) as
\be Z_N=
\prod_{n=0}^{\infty} y_{N_n}(n),\quad\quad 0\le N_n,\quad \quad \sum_nN_n=N
\ee
has, using (\ref{8}) to all orders, the simple factorized form
\be\label{Z}
Z_N= \sum_{n_1,\ldots ,n_N=0}^{\infty} P_{n_1,n_2}(\alpha)P_{n_2,n_3}(\alpha)\ldots
P_{n_{N-1},n_N}(\alpha)e^{-\beta(\epsilon_o(n_1)+
\epsilon_o(n_2)+\ldots +\epsilon_o(n_N))}
\ee
In particular in the $2$-body case 
\be\label{onemore} Z_2-Z_2|_{Fermi}= \sum_{n_1,n_2=0}^{\infty}\left(
P_{n_1,n_2}(\alpha)-P_{n_1,n_2}(\alpha=1)\right)
e^{-\beta(\epsilon_o(n_1)+
\epsilon_o(n_2))}\ee
and thus in the thermodynamic limit\footnote{By factorizing the center of mass,   
(\ref{onemore}) rewrites as
\be Z_2-Z_2|_{Fermi}= {1\over 2\sinh{\beta\omega\over 2}}
\sum_{{\tilde{n}}_2=2l\ge 0} 
\left(e^{-\beta\omega(
{\tilde{n}}_2+{1\over 2}+\alpha)}-e^{-\beta\omega(
{\tilde{n}}_2+{1\over 2}+1)}\right)\ee
and (\ref{twomore}) as
\be Z_2-Z_2|_{Fermi}=({L\over 2\pi})^2\int_{-\infty}^{\infty}dK e^{-\beta
{K^2\over 4}} \int_{-\infty}^{\infty}dk\Phi(2k)e^{-\beta k^2}
\ee
Since,  in the thermodynamic limit \cite{dada} for the $N$-th
cluster coefficient, 
 ${1\over\beta\omega}\to {L\over \sqrt{2\pi\beta}}\sqrt{N}$, 
one infers that in the 2-body case
\be{1\over 2\sinh{\beta\omega\over 2}}\to 
 {L\over 2\pi}\int_{-\infty}^{\infty}dKe^{-\beta {K^2\over 4}}\ee
and therefore one should have
\be\label{10}
\sum_{{\tilde{n}}_2=2l\ge 0} 
\left(e^{-\beta\omega(
{\tilde{n}}_2+{1\over 2}+\alpha)}-e^{-\beta\omega(
{\tilde{n}}_2+{1\over 2}+1)}\right)\to
{L\over 2\pi}
\int_{-\infty}^{\infty} dk
\Phi(2k)
e^{-\beta{k^2}}
\ee
a result that can be trivially checked by direct computation.}
 $\omega\to 0$
\be\label{twomore} Z_2-Z_2|_{Fermi}= ({L\over 2\pi})^2
\int_{-\infty}^{\infty} dk_1dk_2
\Phi(k_1-k_2)
e^{-\beta({k_1^2\over 2}+{k_2^2\over 2})}\ee
where $\Phi$
is given in (\ref{3}).

As far as the LLL-anyon model in a harmonic well \cite{ouvry2} is concerned,
one finds that (\ref{5},\ref{6},\ref{9}) are unchanged,
to the exception of the 1-body
energy which now reads $\epsilon_o(n)=(\omega_t-\omega_c)n+\omega_c$, where
$\omega_c=eB/2$, $\omega_t=\sqrt{\omega_c^2+\omega^2}$, and the
statistical anyonic parameter has to be understood as being $-\alpha$,
i.e. the screening regime where the flux $\phi=-\alpha\phi_o$ ($\phi_o$ is the
quantum of flux) carried by each
anyon is
antiparallel to the external magnetic field.

One can easily convince oneself that in the thermodynamic limit $\omega\to
0$, both the Calogero and LLL-anyon TBA thermodynamics narrow down to
\be\label{a} \log Z=\int_0^{\infty} d\epsilon_o \rho_o(\epsilon_o)\ln (1+ze^{-\beta
\epsilon(\epsilon_o)})\ee
where the dressed energy $\epsilon(\epsilon_o)$ is implicitely 
defined \`a la TBA in terms
of the bare energy $\epsilon_o$ as
\be\label{b}  \beta\epsilon= \beta\epsilon_o
-\int_0^{\infty} d\epsilon' \Phi(\epsilon,\epsilon')\ln (1+ze^{-\beta\epsilon'})\ee
and
\be\label{c} \Phi(\epsilon,\epsilon')=(1-\alpha)\delta(\epsilon-\epsilon')\ee
Here,  $\rho_o(\epsilon_o)$ is
the 1-body
density of states of the bare spectrum of the system considered, i.e.
in the Calogero case the 1d free density of states
$\rho_o(\epsilon_o)= L/(\pi\sqrt{2\epsilon_o})$,
and, in the LLL-anyon case, the 2d LLL density of states,  $\rho_o(\epsilon_o)=
(BV/\phi_o)\delta(\epsilon_o-\omega_c)$, where $V$ is now the infinite surface
of the 2d plane.

In the case of  the Calogero model in a periodic box -the
Calogero-Sutherland model-,
one  can still propose (\ref{5}) and (\ref{6}), but now the
1-body dressed
energy $\epsilon({n})$ should  be defined in terms of the
1-body spectrum in a 1d periodic box of length $L$, $\epsilon_o(n)=k^2/2$, with discretized momentum
$k=2\pi n/L$ and $n$ positive, null or negative integer.

 However, and contrary to the harmonic case, the very fact that the
bare quantum numbers in a periodic box can be of both signs lead to some 
adjustments. If one looks at the $N$-body  Calogero-Sutherland spectrum\footnote{with a BA spectrum
${\bf n}'_i= {\bf n_i}+{\alpha\over 2}\sum_{j\ne i}sign({\bf n'}_i-{\bf
n'}_j)$}
\be \label{19}E_N={1\over 2}({2\pi\over L})^2\sum_{i=1}^N\left([{\bf n}_i+\alpha
(i-1)]-\alpha{(N-1)\over 2}\right)^2\ee
where $ {\bf n}_1\le {\bf n}_2\le \ldots $, one finds that in terms of the bare quantum 
numbers
 $n_i\to {\bf n}_i+\alpha(i-1)-\alpha(N-1)/2= n_{i-1}+\tilde{n}_i+\alpha(i-1)-\alpha(N-1)/2 $ with $
\tilde{n}_i\ge 0$.
This is quite similar to the Calogero-moser spectrum, up to a global shift,
${ n}_i\to { n}_i -\alpha(N-1)/2$, a $N$-dependant periodic
boundary condition adjustment
insuring that the dressed spectrum  remain symmetric
around $0$ in order to minimize the $N$-body energy. 
In the 2-body case  $n_1\to n_1-\alpha/2, n_2\to
n_1+\tilde{n}_2+\alpha/2$, it amounts to the the action of the
Calogero-Moser projector $P_{n_1,n_2}(\alpha)$ as given in (\ref{9})
on the a priori two independent quantum
numbers $ n_1,n_2$, again up to the $2$-body shift
$n_{1,2}\to n_{1,2} -\alpha/2$.
This being considered, altogether with the fact that
the Calogero-Moser and Calogero-Sutherland models originate from the same
model, up to a long distance
regularisation,
it is  natural to take for both  models 
the same TBA function (\ref{9})
to obtain, in view of (\ref{Z}), the correct
Calogero-Sutherland $N$-body partition function, but in addition 
the  shift $n_i\to n_i -\alpha(N-1)/2$ has to be made a posteriori.

At this point, one  can remark that in all  cases  studied so far,
the   Calogero-Moser model, as well as the
Calogero -Sutherland model up to
periodic boundary conditions adjustments, and their thermodynamic limit, the
Calogero model,
the TBA functions $\Phi_{n_1,n_2}$ and $\Phi(k_1-k_2)$
are intimately related to the relative $2$-boson density of states for the
problem at hand.
Indeed, in a harmonic well, the  spectrum for a relative particle 
with bosonic
statistics and 
 interacting with a Calogero potential at the origin is
\be\label{omega} \epsilon=\omega(n+{1\over 2}+\alpha )\quad 0\le n\ee
with  $n$ even, i.e. with symmetric eigenstates under $x\to -x$.

Let us first consider, in the thermodynamic limit, the Calogero model:
when $\omega\to 0$, the relative 2-boson density of
states  reads
\be \rho_{\alpha}(\epsilon)-\rho_{\alpha=1}(\epsilon)= {1-\alpha\over 2}
\delta(\epsilon)\ee
It rewrites in terms of the relative momentum $k$ such that $\epsilon=k^2$
\be \rho_{\alpha}(k)-\rho_{\alpha=1}(k)= {}{1-\alpha\over 2}
\delta(k)\ee
Now one has to map the relative 2-body momentum $k$ on the ``momentum''
$k_2-k_1$ the function $\Phi(k_2-k_1)$ is concerned with. Since
 $k_2-k_1=2k$, one gets for the density of states in
terms of $k_2-k_1$ 
\be
\rho_{\alpha}({k_2-k_1\over 2})-\rho_{\alpha=1}({k_2-k_1\over 2})=(1-\alpha)\delta(k_2-k_1)\ee
i.e. precisely (\ref{3}) up to a factor $2\pi/L$.

When $\omega$ is kept finite, the same logic applies:
the relative spectrum (\ref{omega}) yields
\be\label{rel} n\to n+\alpha\ee
One has yet to map the relative 2-body bosonic even quantum
number $n$ on the ``quantum number'' $n_2-n_1$
that the function
$\Phi_{n_1,n_2}$ is concerned with.
For a given 2-body energy, i.e. for ${\bf n}_1+{\bf n}_2$ given, one has
${\bf n}_2-{\bf n}_1=n$ -then the center of mass quantum number is $2 {\bf
n}_1$, or ${\bf n}_2-{\bf n}_1=n+1$ -then the center of mass quantum number is
$2{\bf n}_1+1$, depending 
if ${\bf n}_2-{\bf n}_1$  is even or odd. 
One finds that $n\to n+\alpha$ rewrites, in terms of the independent
$n_1,n_2$ as
$n_1\to n_1$, $n_2\to n_1+{\tilde {n}_2}+\alpha$, where now ${\tilde {n}_2}=n$
or ${\tilde {n}_2}=n+1$, i.e. any positive integer. Then (\ref{rel})
is indeed identical to the action of $\Phi_{n_1,n_2}$
in (\ref{9}).

This is not a surprise, scattering 2-body phase shifts are known to be
linked to the $2$-body density of states via
S-matrix arguments \cite{smatrix}. 

\section{The Lieb-Liniger case}

In the Lieb-Liniger model in the
thermodynamic limit,
the same conclusion happens to be  true.
The model, defined as 
\be H_N=-{1\over 2}\sum_{i=1}^N{d^2\over dx_i^2}+c\sum_{i<j}\delta(x_i-x_j)\ee
is solvable by Bethe ansatz \cite{LL} and has a TBA thermodynamics \cite{YY} 
obtained from
\be\label{YY} \Phi(k_1-k_2)={1\over L}{2c\over (k_2-k_1)^2+c^2}\ee
It interpolates between the Bose ($c=0$) and Fermi ($c=\infty$)
thermodynamics
{\bf and describes particles
with intermediate statistics \cite{myrheim}.}
For a relative
particle with bosonic
statistics interacting with a $\delta$ potential at the origin the density
of states is 
\be \rho_c(\epsilon)-\rho_{c=\infty}(\epsilon)={}{1\over 4\pi\sqrt
\epsilon}{c\over
\epsilon+{c^2\over 4}}\ee
which in terms of $\epsilon=k^2$, $k>0$ rewrites as
\be \rho_c(k)-\rho_{c=\infty}(k)={}{1\over 2\pi}{c\over
k^2+{c^2\over 4}}\ee
Now, one has again to map  $k$ on the ``momentum''
$k_2-k_1$ the function $\Phi(k_1,k_2)$ is concerned with, i.e. 
$k_2-k_1=2k$, and since $k_2-k_1$ can be either positive or negative, one gets
for the density of states in terms of $k_2-k_1$
\be \rho_c({k_2-k_1\over 2})-\rho_{c=\infty}({k_2-k_1\over 2})={1\over 2\pi}{2c\over
(k_2-k_1)^2+{c^2}}\ee
i.e. nothing but (\ref{YY}), again up to a factor $2\pi/L$.

If one follows the same line of reasoning which
was operative in the Calogero-Moser case
to obtain the discrete TBA function $\Phi_{n_1,n_2}$  (\ref{9}) 
from the relative
2-body spectrum (\ref{omega}),
one might try,  for  the Lieb-Liniger model in a 
harmonic well,  
 discrete TBA
thermodynamics (\ref{5},\ref{6}) equations  
with a TBA function
$\Phi_{n_1,n_2}$  deduced  from the $2$-body 
relative bosonic
spectrum in a harmonic well \cite{myrheim}.
 It rewrites as
 \be \label{myr} \epsilon=\omega\left(n+{1\over 2}+{2\over \pi}\arctan \left( {c\over
2\sqrt{2\omega}}{\Gamma({\epsilon\over 2\omega}+{1\over 4})\over
\Gamma({\epsilon\over 2\omega}+{3\over 4})}\right)\right)\quad 0\le n\ee
with $n$ even, i.e. as
$\epsilon=\omega(n+{1\over 2}+f_c(n))$
with{\bf  \footnote{Equivalently,  starting from the Fermi spectrum 
by rewriting $f_c(n)=1-g_c(n)$
\be\label{interr} 0\le g_c(n)={2\over \pi}\arctan \left( {2\sqrt{2\omega}\over
c}
{\Gamma({n+3\over 2}-{g_c(n)\over 2})\over\Gamma(
{n+2\over 2}-{g_c(n)\over 2})}\right)\le 1\ee
}}
\be\label{inter} 0\le f_c(n)={2\over \pi}\arctan \left( {c\over
2\sqrt{2\omega}}{\Gamma({n+1\over 2}+{f_c(n)\over 2})\over
\Gamma({n+2\over 2}+{f_c(n)\over 2})}\right)\le 1
\ee
and  interpolates  between the relative $2$-body bosonic
($c=0$, $f_0(n)=0$, $g_0(n)=1$) and fermionic ($c=\infty$, $f_{\infty}(n)=1$, 
$g_{\infty}(n)=0$) spectra.

Therefore let us try for the    Lieb and
Liniger 
in a harmonic well 
the discrete TBA function
\be \label{endd}\Phi_{n_1,n_2}=P_{n_1,n_2}(c)-P_{n_1,n_2}(c=\infty)\ee
should be  defined in terms of $P_{n_1,n_2}(c)$ such that, $n_1$ being 
left unchanged,
\be\label{cc} n_2\to n_1+{\tilde{n}}_2+f_c({\tilde{n}}_2)\ee
if ${\tilde{n}}_2\ge 0$ is even, and
\be\label{ccc} n_2\to n_1+{\tilde{n}}_2+f_c({\tilde{n}}_2-1)\ee
if ${\tilde{n}}_2$ is odd.
 Note again that substracting $P_{n_1,n_2}(c=\infty)$ in (\ref{endd}) originates,
as in the Calogero case,
from the fermionic convention (obviously 
$P_{n_1,n_2}(c=\infty)=P_{n_1,n_2}(\alpha=1)$.)

 It is easy to check that 
the $2$-body partition function is
reproduced by the discrete TBA equations
\be\label{c'est} Z_2-Z_2|_{Fermi}=\sum_{n_1=n_2=0}^{\infty}( P_{n_1,n_2}(c)-
P_{n_1,n_2}(c=\infty) )
e^{-\beta(\epsilon_o(n_1)+\epsilon_o(n_2))}
\ee
and thus, 
in the thermodynamic limit ${\omega\to 0}$,
\be\label{arcs}
\sum_{{\tilde{n}}_2=2l\ge 0} 
\left(e^{-\beta(
{\tilde{n}}_2+{1\over 2}+f_c({\tilde{n}}_2))}-(c=\infty)\right)
\to
{L\over 2\pi}\int_{-\infty}^\infty dk
\Phi(2k)
e^{-\beta k^2}\ee
where the function $\Phi$ is given in (\ref{YY}),
a result  that can be checked by direct computation, order by order in $1/c$.
There are two independant dimensionless parameters, $\beta\omega$
(thermodynamic limit)
and $\sqrt{\beta} c$ (``coupling constant'').
Clearly, for a given
coupling constant   $\sqrt{\beta} c$,
looking at $f_c(n) =1-g_c(n)$ in (\ref{inter},\ref{interr}),
one has to consider, in the thermodynamic limit $\beta\omega\to 0$,
the spectrum close to the Fermi point $(c=\infty)$,  
\be\label{hole} g_c(2l) = 4{\sqrt{2}}{\sqrt{\beta\omega}\over \pi \sqrt{\beta} c}
{\Gamma(l+{3\over2})\over l!}+\ldots \ee
from which  (\ref{arcs}) can be recovered, here at first order $1/(\sqrt{\beta}c)$.

Note also that in the thermodynamic limit for the relative spectrum, with
 $(n+1/2)\omega\to k^2$, i.e. $n\omega$  fixed,
(\ref{inter}) becomes
\be f_c(k)={2\over \pi}\arctan {c\over 2k}\ee
which is indeed reminiscent of the Lieb and Liniger BA
spectrum \cite{LL}.

What about the $N$-body problem? 
A possible way to check the discrete TBA is to see if 
the perturbative TBA
thermodynamics coincide with the exact
(standard) Hamiltonian perturbative thermodynamics \cite{dasnieres}, which can be
computed  with the Lieb
and Liniger Hamiltonian  from the Bose point $c=0$ (from the  Fermi point
$c=\infty$ standard perturbation theory is   meaningless).
Perturbation theory yields
\be\label{alix} \log Z= \log Z|_{Bose}+\sqrt{\beta}c
\sum_{s,t=1}^{\infty}{z^{s+t}\over4\sqrt{\pi}} {\sqrt{\beta\omega}\over
\sqrt{\sinh{s\beta\omega\over 2}
\sinh{t\beta\omega\over 2}\sinh{(s+t)\beta\omega\over 2}}}+\ldots \ee
Let us now  consider the $\sqrt{\beta} c$ expansion
from the TBA point of view, again
from the Bose point. One has to consider
(\ref{inter}) at first order in $\sqrt{\beta}c$ 
\be\label{wrong} f_c(2l)={1\over \pi\sqrt{2}}{\sqrt{\beta}c\over \sqrt{\beta\omega}}
{\Gamma(l+{1\over 2})\over l!}+\ldots \ee
 and compute from the discrete TBA equations 
\be\label{martin} \log Z= \log Z|_{Bose}+
\sum_{n_1=0}^{\infty}{ze^{-\beta\omega(n_1+{1\over 2})}\over 1-
ze^{-\beta\omega(n_1+{1\over 2})}
}\sum_{n_2=0}^{\infty}\Phi_{n_1,n_2}\ln{1\over
1-ze^{-\beta\omega(n_2+{1\over 2})}}|_1+\ldots 
\ee
where $\Phi_{n_1,n_2}\ln{1\over
1-ze^{-\beta\omega(n_2+{1\over 2})}}|_1$ means evaluating this expression at first order
in $\sqrt{\beta}c$ using (\ref{wrong}).
One finds
\be \label{non} \log Z= \log Z|_{Bose}+\sqrt{\beta}c
\sum_{s,t=1}^{\infty}{z^{s+t}\over 4\sqrt{2\pi}}{\sqrt{\beta\omega}\over
\sinh{(s+t)\beta\omega\over 2}}
\left(\sqrt{\coth{s\beta\omega\over 2}}+
      \sqrt{\coth{t\beta\omega\over 2}}\right)+\ldots \ee
which   coincides with (\ref{alix})  only in the limit
$\beta\omega\to \infty$, i.e. for a given $\omega$, in the zero temperature
limit\footnote{The limit $\omega\to\infty$ for a given temperature is not
considered here. In this limit all particles are confined at $x_i=0$. But
$\delta$ interactions actually forbid this unless the effective coupling
constant vanishes, which is precisely happening  in the 2-body
case (\ref{inter}). In other words  the $\omega\to \infty$ limit
is the trivial bosonic limit.}, i.e the groundstate.

In fact, discrete TBA gives in the
vanishing temperature  limit direct information on
the $N$-body groundstate energy : in the Calogero-Moser case, it is obtained,
in the bosonic based formulation,
by restricting the discrete TBA equations

\be
\log Z=
\sum_{n_1=0}^{\infty}
\log{1\over 1-ze^{-\beta{\tilde{\epsilon}}(n_1)}},
\ee

\be
\beta{\tilde{\epsilon}}({n_1})=\beta \epsilon_o(n_1)
-\sum_{n_2=0}^{\infty}
{\tilde{\Phi}}_{n_1,n_2}
\log{1\over 1-ze^{-\beta{\tilde{\epsilon}}({n_2})}}
\ee
with ${\tilde{\Phi}}_{n_1,n_2}= P_{n_1,n_2}(\alpha)- P_{n_1,n_2}(\alpha=0)$
to the groundstate quantum numbers
$n_1=0$ and ${\tilde{n}_2}=0$. One obtains 
\be E_N^{G.S.}=\omega\left({N\over 2}+ {N(N-1)\over 2}\alpha\right)\ee
In the Lieb and Liniger case a similar approach gives 
\be E_N^{ G.S.}=\omega\left({N\over 2}+{N(N-1)\over 2}f_c(0)\right)\ee
a result which is consistent with the  groundstate
energy at first order in $c$ 
\be \label{paris}E^{G.S.}_N=\omega\left({N\over 2}+{N(N-1)\over
2}f_c^{(1)}(0)c+\ldots\right)\ee
where $f_c^{(1)}(0)=1/\sqrt{2\pi\omega}$ stands for the first order term in the
expansion of $f_c(0)$ in power of $c$.
However, second order standard perturbation theory
gives corrections to the Calogero-Moser like energy  $\omega N(N-1)f_c(0)/2 $.
For example in the $N=3$ case 
\be \label{pariss}E^{G.S.}_3=\omega\left({3\over 2}+ 3f_c^{(1)}(0)c+ 3\left(f_c^{(2)}(0)-{1\over \pi\omega}\log{4\over
2+\sqrt{3}}\right)c^2+\ldots\right) 
\ee
where $f_c^{(2)}(0)=-{\log 2\over 2\pi\omega}$ stands for the second order
term in the expansion of $f_c(0)$.

\section{Conclusion}
We have shown how the Caloger-Moser thermodynamics can be
 rewritten in terms of discrete TBA equations. In the Calogero-Sutherland model,
 the same TBA equations were shown to be operative, up to a global shift of the bosonic
 quantum numbers. Since the  Lieb-Liniger model 
 shares common features with the
 Calogero model -BA solvability, TBA thermodynamics in the
 thermodynamic limit, intermediate statistics- it might
also  have, when considered in a harmonic well, a discrete TBA thermodynamics.
We tried to illustrate this point of view by proposing discrete TBA equations for
the harmonic Lieb and Liniger model in analogy with
the Calogero-Moser TBA thermodynamics.
However the groundstate energy shows deviations from this TBA framework at
second order in perturbation theory. 

We leave to a further study to find   analytical or
 numerical ways  to improve and give a stronger basis to
 the  discrete  TBA thermodynamics for the Lieb-Liniger model, and in particular   extract a useful information 
on the groundstate for a given density of particles.
It would however certainly be  interesting to understand
 more in detail the zero temperature limit of a system which is supposed to
 describe the physics of 1d Bose Einstein condensates in harmonic traps.

Acknowledgements: One of us (S.O.) would like to thank S. Matvenkoo for
useful discussions.

\end{document}